# Epitaxial Integration of a Perpendicularly Magnetized Ferrimagnetic Metal on a Ferroelectric oxide for Electric-Field Control

Xin Zhang, Pei-Xin Qin, Ze-Xin Feng, Han Yan, Xiao-Ning Wang, Xiao-Rong Zhou, Hao-Jiang Wu, Hong-Yu Chen, Zi-Ang Meng, Zhi-Qi Liu*

*School of Materials Science and Engineering, Beihang University, Beijing 100191, China*
Email: zhiqi@buaa.edu.cn

**Abstract:** Ferrimagnets, which contain the advantages of both ferromagnets (detectable moments) and antiferromagnets (ultrafast spin dynamics), have recently attracted great attention. Here we report the optimization of epitaxial growth of a tetragonal perpendicularly magnetized ferrimagnet $Mn_2Ga$ on MgO. Electrical transport, magnetic properties and the anomalous Hall effect (AHE) were systematically studied. Furthermore, we successfully integrated high-quality epitaxial ferrimagnetic $Mn_2Ga$ thin films onto ferroelectric PMN-PT single crystals with a MgO buffer layer. It was found that the AHE of such a ferrimagnet can be effectively modulated by a small electric field over a large temperature range in a nonvolatile manner. This work thus demonstrates the great potential of ferrimagnets for developing high-density and low-power spintronic devices.

# 1. Introduction

Contemporary mass storage for data centers predominantly relies on hard disk drives that are based on perpendicularly magnetized ferromagnetic granular films, the spin states of which can be easily manipulated by external magnetic fields generated by current coils due to remarkable macroscopic moments. However, as limited by the characteristic GHz spin dynamics, ferromagnetic materials could hardly be utilized for the static random-access memory technology and the in-memory computing, which needs a sub-ns response speed. Similar to ferromagnets, ferrimagnets possess large magnetic moments. Besides, they exhibit ultrahigh spin dynamics of THz as a result of antiferromagnetic exchange coupling [1-6] akin to antiferromagnets. Hence, ferrimagnets are rising star materials for new-generation sub-ns information devices [7-12].

Tetragonal $D0_{22}$ $Mn_2Ga$ with $a$ = 3.905 Å and $c$ = 7.193 Å, a classical ferrimagnet with a high Curie temperature of ~710 K, exhibits large magnetic anisotropy along its [001] crystallographic orientation. (001)-oriented single-crystalline or textured films are thus perpendicularly magnetized. In addition, $D0_{22}$ $Mn_2Ga$ is of high spin polarization at the Fermi level and low Gilbert damping constant [13], both of which are useful for spin valves, perpendicular magnetic tunnel junctions [14], narrow-band terahertz emission from coherently excited spin precession [15], and high-density spin-transfer-torque magnetoresistive random access memories (STT-MRAM) [16].

However, the information writing (corresponding to the spin state manipulation) of $Mn_2Ga$-based spintronic devices such as spin valves and STT-MRAM mostly depend on electrical-current-generated magnetic fields or electrical currents, which generates significant Joule heating and hence results in high power consumption. Alternatively, if one can effectively control the spin state of $Mn_2Ga$ with a non-electrical-current manner, the energy for writing a bit could be substantially lowered.

An electric field applied onto a conductor yields an electrical current. In contrast, for highly insulating ferroelectric oxide materials, the application of an electric field generates a negligible current. Instead, it induces piezoelectric strain [1-7,17-23]. Assuming (001)-oriented $D0_{22}$ $Mn_2Ga$ could be epitaxially integrated onto ferroelectric oxides, electric-field-generated strain could harness the spin state of Mn$_2$Ga as the electronic states of solid-state materials are all sensitive to the periodic lattice.

Nevertheless, epitaxial growth of intermetallic Mn$_2$Ga on a ferroelectric oxide is quite challenging. The key reasons are: (1) highly-ordered epitaxial films need high thermal energy to let the atoms diffuse into their equilibrium sites during thin film growth and therefore high growth temperatures are required; (2) intermetallic alloys are of strong chemical activity, which can easily be oxidized or form secondary alloys at high temperatures while landing on ferroelectric oxide substrates; (3) lattice mismatch between intermetallic alloys and ferroelectric oxides could break epitaxial growth. Accordingly, a proper oxide substrate or an oxide buffer layer with robust high-temperature stability and low oxygen diffusion coefficient could be crucial for realizing epitaxial growth of Mn$_2$Ga on ferroelectrics. In this letter, we report the optimal epitaxial growth of D0$_{22}$ Mn$_2$Ga on ferroelectric PMN-PT. Furthermore, electric-field control of its AHE has been achieved over a large temperature range, which paves the way for low-power Mn$_2$Ga spintronic device applications.

## 2. Experimental

MgO, a transparent insulating oxide with a large bandgap of 7.8 eV, shows excellent high-temperature stability with a melting point of 2800°C. It has a cubic lattice with $a$ = 4.212 Å, which is close to the in-plane lattice constant of D0$_{22}$ Mn$_2$Ga. Therefore, Mn$_2$Ga films were firstly grown on single-crystalline (001)-oriented MgO substrates using the magnetron sputtering technique at different temperatures ranging from 30 to 550°C. The base pressure and the D.C. sputtering power were 7.5×10$^{-9}$ Torr and 60 W, respectively. During the deposition, the Ar pressure was kept at 3mTorr. With X-ray reflectometry, the deposition rate was determined to be 35 Å/min and the total thickness was first kept at 100 nm, which was later changed to 50 and 30 nm for thickness dependence studies. The crystal structure of Mn$_2$Ga thin films was measured by X-ray diffraction (XRD). Magnetic characterization and electrical measurements were performed by a Quantum Design VersaLab with a vibrating sample magnetometer option. The standard linear four-probe method and the Hall geometry were used for longitudinal and Hall resistance measurements, respectively.

## 3. Results and discussion

Figure 1a shows single-crystal XRD spectra of $Mn_2Ga$ films grown on MgO substrates fabricated at different growth temperatures $T_G$. For $T_G$ below 450°C, no thin-film peaks are seen, implying disordered polycrystalline films. At 450°C, the (002) peak and a weak (001) peak of $D0_{22}$ $Mn_2Ga$ show up. With increasing the growth temperature to 550°C, both the (001) and (002) peaks of $D0_{22}$ $Mn_2Ga$ become sharper, indicating enhanced crystallinity and chemical ordering. The growth temperature could not be further raised. That is because at higher temperatures the surface energies of intermetallic $Mn_2Ga$ and ferroelectric oxide PMN-PT are of large different, due to the non-wetting issue, $Mn_2Ga$ could not form continuous thin films but only separate islands.

The metallicity, which could be examined by the normalized resistivity relative to room-temperature values, is demonstrated in Fig. 1b. When the substrate temperature is lower than 450°C, $Mn_2Ga$/MgO films are semiconducting, which is contrary to its bulk metallic behavior. This suggests the existence of a large degree of chemical disorder including grain boundaries. The films fabricated at 450 and 550°C are metallic and the metallicity is improved with enhancing substrate temperature. All these electrical transport results are consistent with the XRD spectra shown in Fig. 1a.

The out-of-plane and in-plane magnetic moments versus magnetic fields (*M-H*) are measured at 50 and 300 K for single-crystalline $Mn_2Ga$/MgO films (Fig. 1c, f). Overall, both films exhibit the feature of perpendicularly magnetized anisotropy. The room-temperature saturation magnetization of $Mn_2Ga$/MgO films are ~290 and ~335 emu/cc for growth temperature of 450 and 550°C, respectively, which are comparable with the magnetization values of previously reported ferrimagnetic $Mn_2Ga$ films [24]. Similar to the XRD and electrical resistivity data, the increase of the growth temperature improves the squareness of the out-of-plane *M-H* loop, which is favorable for perpendicular spintronic device applications. Therefore, 550°C is the optimized growth temperature for high-quality epitaxial and perpendicularly magnetized ferrimagnetic $Mn_2Ga$ films.

In addition, the room-temperature anisotropy field $\mu_0 H_k$ is determined to be ~10 T for the 550°C-fabricated $Mn_2Ga$ film, which corresponds to a uniaxial magnetocrystalline anisotropy energy $K_u$ of ~1.68 MJ/m$^3$, which is of the same order with the previously reported highest $K_u$ (~2.17 MJ/m$^3$) [25] for molecular-beam-epitaxy-fabricated ferrimagnetic Mn-Ga films. For real applications, the duration of information storage needs to be more than 10 years, which requires the ratio of magnetocrystalline energy of a bit $K_uV$ greater than 40 times room-temperature thermal energy $40k_BT$ ~ 1.67×10$^{-19}$ J. Considering a bit cell consisting of a single Mn$_2$Ga layer with a cubic shape, the critical bit size could accordingly be reduced to ~4.6 nm. This thus implies great potential of our optimized Mn$_2$Ga ferrimagnetic thin films with perpendicular magnetized anisotropy for high-density data storage.

Longitudinal magnetotransport properties of Mn$_2$Ga thin films fabricated at various temperatures were examined for out-of-plane magnetic fields. It was found that for $T_G$ < 300°C, the magnetoresistance (MR) effect is rather weak and almost not affected by the growth temperature. Figure 2 shows the MR curves collected at different temperatures ranging from 50 to 300 K for Mn$_2$Ga thin films with $T_G$ > 30°C. For $T_G$ = 150°C (Fig. 2a), the MR above 50 K is positive, which implies that the orbital scattering due to the Lorentz force is dominant. However, the MR turns into negative for 50 K, suggesting the important role of magnetic moments of Mn. For $T_G$ = 300°C (Fig. 2b), the room-temperature MR is negligible while the low-temperature MR curves are interestingly linear. It is worth noticing that the positive MR at 150 K is the largest, reminiscent of the maximal magnetotransport properties at ~200 K for noncollinear antiferromagnets Mn$_3$Sn [26] and Mn$_3$Ge [4]. For crystallized epitaxial Mn$_2$Ga films, the butterfly-shape hysteresis MR curves (Fig. 2c, d) are clearly seen, characteristic of long-range ferrimagnetic/ferromagnetic order. In addition, the positive orbital scattering is more significant at low temperatures, leading to suppressed negative MR.

Systematic transverse magnetotransport properties, *i.e.*, the Hall effect, of the Mn$_2$Ga/MgO films deposited below 450°C are demonstrated in Fig. 3. Similarly, the Hall curves for $T_G$ = 30 and 150°C (Fig. 3a, b) are comparable and are linear above 50 K. At 50 K, the Hall effect becomes nonlinear, signature of the magnetic-moment-related AHE, which is consistent with the negative MR at 50 K in Fig. 2a. For $T_G$ = 300°C, the AHE is obvious below 200 K, indicating the formation of magnetic order.

The Hall effect of epitaxial $Mn_2Ga$ films is shown in Fig. 4a, b. The general shape of the Hall curves is in excellent agreement with that of the *M-H* loops. Therefore, the Hall effect could serve as a sensitive electrical probe to magnetic properties of perpendicularly magnetized $Mn_2Ga$. Detailed scaling law analysis [7] (Fig. 4c) on the optimized film reveals that for low longitudinal resistivity range $280 < \rho_{xx} < 330$ μΩ·cm, $\rho_{xy} \propto \rho_{xx}^2$, the Berry curvature is the dominant origin for the AHE, which is a pseudo magnetic field in momentum space and determined by the topological bands interaction of Bloch electrons [27,28]. While for the large resistivity region with $\rho_{xx} > 330$ μΩ·cm, skew scattering becomes more important for generating a transverse Hall voltage, leading to $\rho_{xy} \propto \rho_{xx}$ [29]. The excellent perpendicular magnetic anisotropy remains in thinner films such as 50 and 30 nm (Fig. 5), which, in turn, lead to significantly enhanced anomalous Hall resistance. The much larger anomalous Hall resistance in thinner $Mn_2Ga$ films could facilitate the electrical read-out for memory devices.

Based on the experimental results mentioned above, 30-nm-thick $Mn_2Ga$ films were further epitaxially integrated onto (001)-oriented PMN-PT ferroelectric oxides with a 25-nm-thick MgO buffer layer so as to manipulate its AHE or magnetism by electric-field-induced piezoelectric strain [1-6,17-23,30]. The MgO buffer layers were grown by a pulsed laser deposition system at 400°C and post-annealed at 600°C for 1 h, which was utilized to prevent Pb element in PMN-PT substrates from diffusing into the chamber and $Mn_2Ga$ films to form secondary alloys at high temperatures [18,31]. As shown in Fig. 6a, the XRD spectrum containing the (002) peak of the MgO buffer layer and the (001) and (002) peaks of the $Mn_2Ga$ thin film indicates the epitaxial growth of the $Mn_2Ga$ on the MgO buffer layer. The field-dependent Hall signals of an epitaxial $Mn_2Ga$/MgO/PMN-PT heterostructure (Fig. 6b) at different temperatures are in consistence with that of $Mn_2Ga$/MgO in Fig. 4b, which suggests the excellent perpendicular magnetic anisotropy of the epitaxially integrated $Mn_2Ga$ films on ferroelectric PMN-PT.

To explore the effect of piezoelectric strain on the AHE in the $Mn_2Ga$/MgO/PMN-PT heterostructure, an electric field $E_G$ of -5 kV/cm was perpendicularly applied across the PMN-PT substrate (Fig. 7a) to pole the ferroelectric substrate at room temperature. To examine any possible variation of the AHE, the Hall curves were re-measured after electric poling of the PMN-PT. It turns out that under such an electric-field-excitation, the AHE is enhanced for all the temperatures (Fig. 7b-g). The relative electric-field-induced nonvolatile modulation of the zero-field anomalous Hall resistance is extracted and plotted in Fig. 7h, which reaches ~16% below 200 K and ~14% at 300 K.

To further confirm the nonvolatile nature of the electric-field-induced piezoelectric strain in PMN-PT, the room-temperature electric-field dependent longitudinal resistance of the $Mn_2Ga$ film in the $Mn_2Ga$/MgO/PMN-PT heterostructure is measured with the linear four-probe geometry (Fig. 8a). As shown in Fig. 8b, the positive and negative peaks in perpendicular gating current through the MgO buffer layer and PMN-PT clearly exhibit the reversible ferroelectric polarization switching feature. Correspondingly, the electric-field-dependent longitudinal resistance (Fig. 8c) shows an asymmetric and nonvolatile butterfly loop, which is similar to what we obtained in previous measurements [3,23].

Empirically, the AHE in ferromagnetic materials is closely related to magnetization. Motivated by this understanding, we examined the out-of-plane magnetization change of the $Mn_2Ga$ film for the $Mn_2Ga$/MgO/PMN-PT heterostructure upon electric-field poling (Fig. 9a) of the ferroelectric substrate PMN-PT. As shown in Fig. 9b, c, the perpendicular magnetization has been changed substantially. At 50 K, the electric-field poling of the PMN-PT alter the saturation magnetization of $Mn_2Ga$ alters from ~360 to ~423 emu/cc (Fig. 9b), which corresponding to an ~17.5% magnetization enhancement, similar to the anomalous Hall resistance increase in Fig. 7b. For 300 K, the out-of-plane magnetization changes from ~335 to ~382 emu/cc (Fig. 9c), well consistent with the ~14% anomalous Hall resistance variation in Fig. 7g. Thus, these experimental results clearly illustrate that the piezoelectric-strain-induced anomalous Hall effect modulation is predominantly caused by the strain-induced magnetization variation. For ferrimagnetic materials with two opposite unequal sublattices, the enlargement of the net magnetization would likely pertain to the weakening of the compensation of two sublattices in terms of the spin rotation, reminiscent of the scenario of noncollinear antiferromagnetic spin structure modulation by piezoelectric strain as theoretically described by Lukashev *et al*. [32].

## 4. Conclusions

In conclusion, we have fabricated epitaxial ferrimagnetic $Mn_2Ga$ thin films with perpendicular magnetic anisotropy on MgO substrates. The mechanisms of the AHE were unveiled for different longitudinal resistivity ranges. When MgO is used as buffer layer, Mn$_2$Ga thin films with perpendicular anisotropy have been successfully integrated onto ferroelectric PMN-PT substrates which is useful for utilizing ferrimagnetic materials in high-density spintronic devices and could enable the fabrication of other exotic epitaxial heterostructures with ferrimagnets interfacing with some novel materials [33-45]. Via the defects engineering in thin films [46], the spin structure and the AHE of ferrimagnetic Mn$_2$Ga could further be modulated to realize the topological Hall effect. More importantly, the AHE of ferrimagnetic Mn$_2$Ga films is largely modulated by the electric-field-induced piezoelectric strain, which paves the way for magnetic-field-free low-power ferrimagnetic spintronic device applications.

**Acknowledgements:** Zhi-Qi Liu acknowledges financial support from National Natural Science Foundation of China (NSFC Grant Nos: 52121001, 51822101, 51861135104 & 51771009).

**Figure 1**

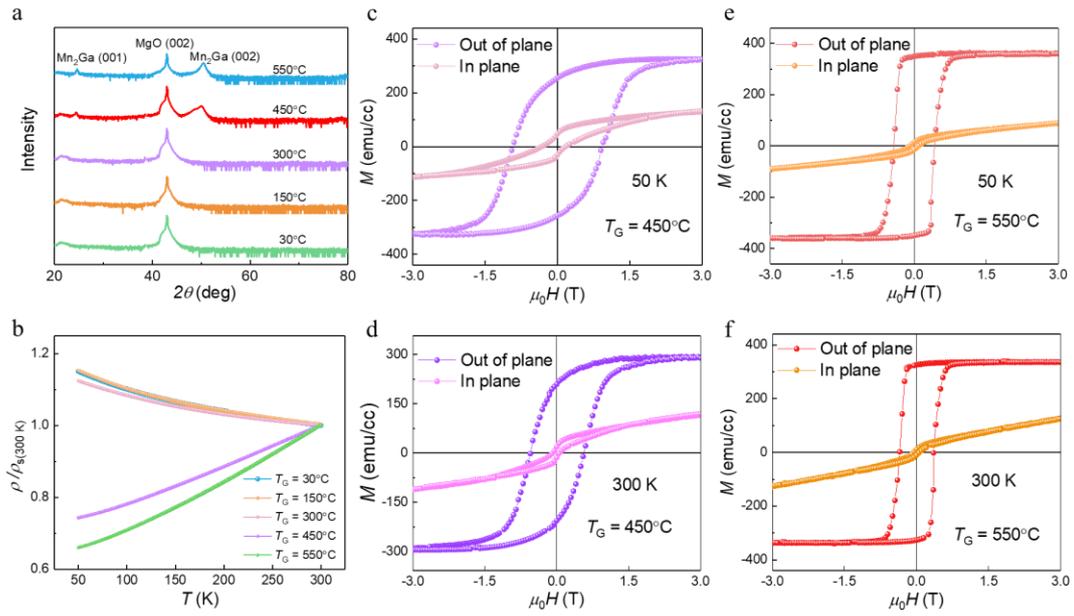

**Fig. 1 a** X-ray diffraction patterns of 100-nm-thick $Mn_2Ga$/MgO heterostructures grown at different temperatures; **b** Temperature-dependent normalized resistivity of $Mn_2Ga$ thin films deposited at different temperatures; **c-f** The magnetization curves of $Mn_2Ga$ samples deposited at 450 and 550°C are measured at 50 and 300 K, respectively

**Figure 2**

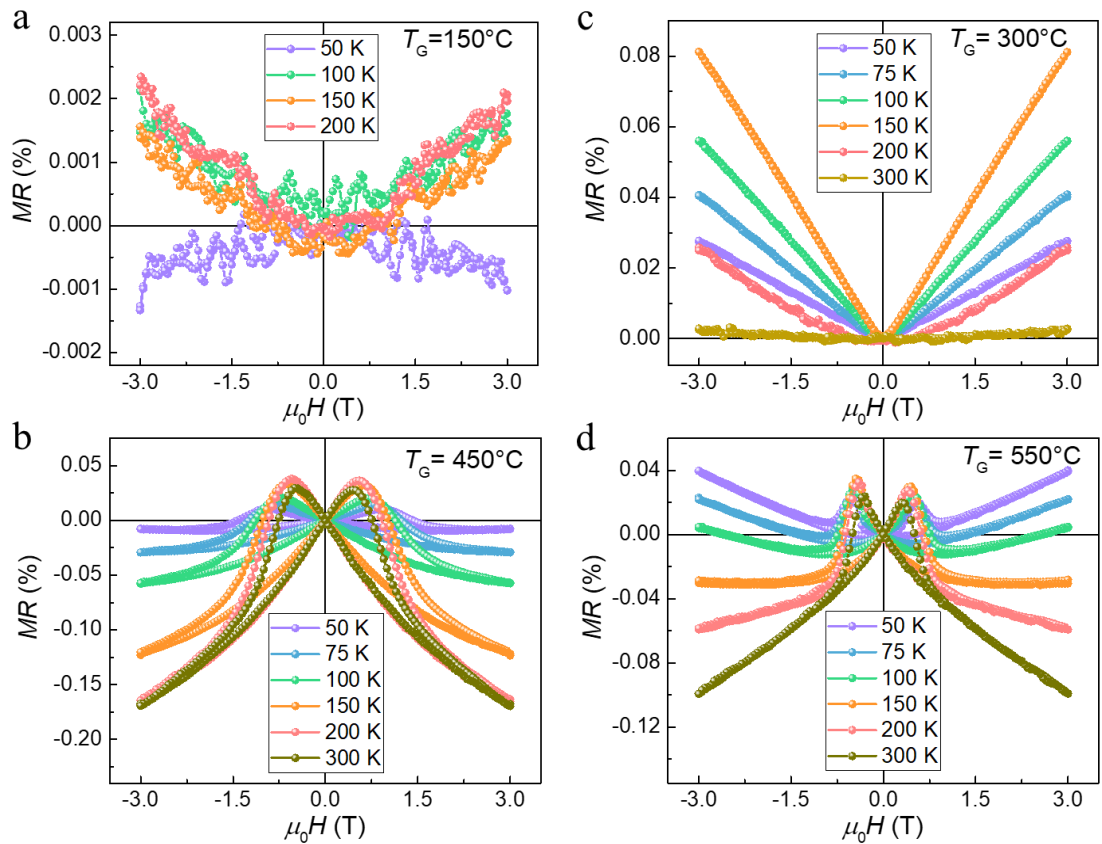

**Fig. 2** Magnetoresistance of 100-nm-thick Mn$_2$Ga/MgO films fabricated at different growth temperatures $T_G$. **a** $T_G$ = 150°C; **b** $T_G$ = 300°C; **c** $T_G$ = 450°C; **f** $T_G$ = 550°C

**Figure 3**

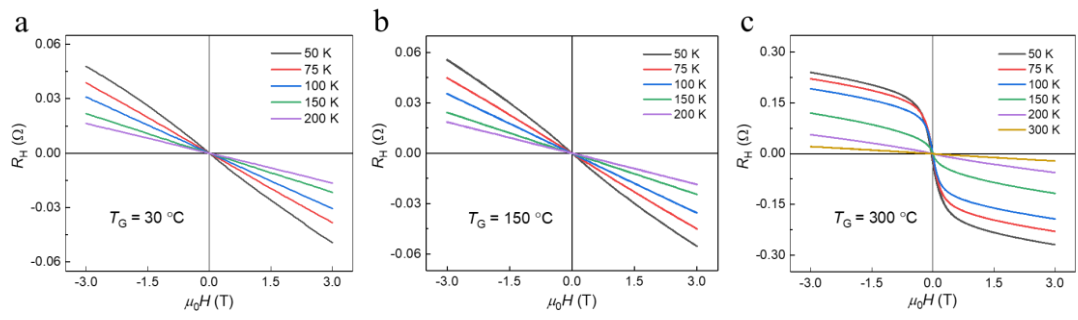

**Fig. 3 Hall effect for 100-nm-thick Mn₂Ga/MgO films fabricated below 450°C. a** $T_G$ = 30°C; **b** $T_G$ = 150°C; **c** $T_G$ = 300°C

**Figure 4**

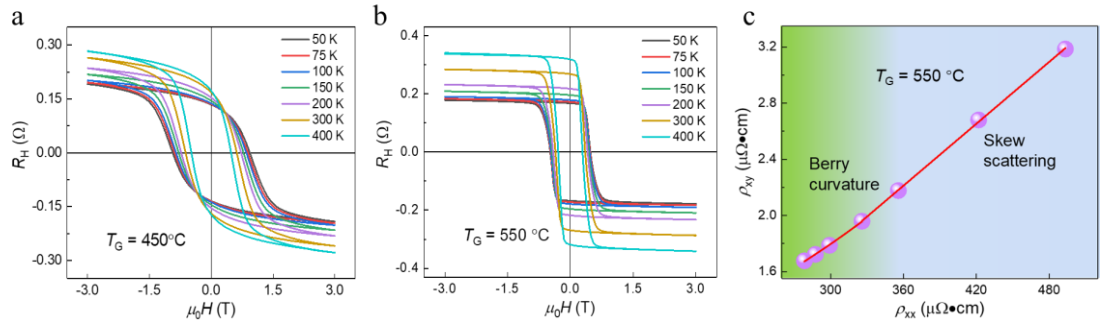

**Fig. 4 a** Hall effect for a 100-nm-thick $Mn_2Ga/MgO$ film fabricated at $T_G$ = 450°C; **b** Hall effect for a 100-nm-thick $Mn_2Ga/MgO$ film fabricated at $T_G$ = 550°C; **c** Scaling law analysis for the $Mn_2Ga$ film fabricated at $T_G$ = 550°C

**Figure 5**

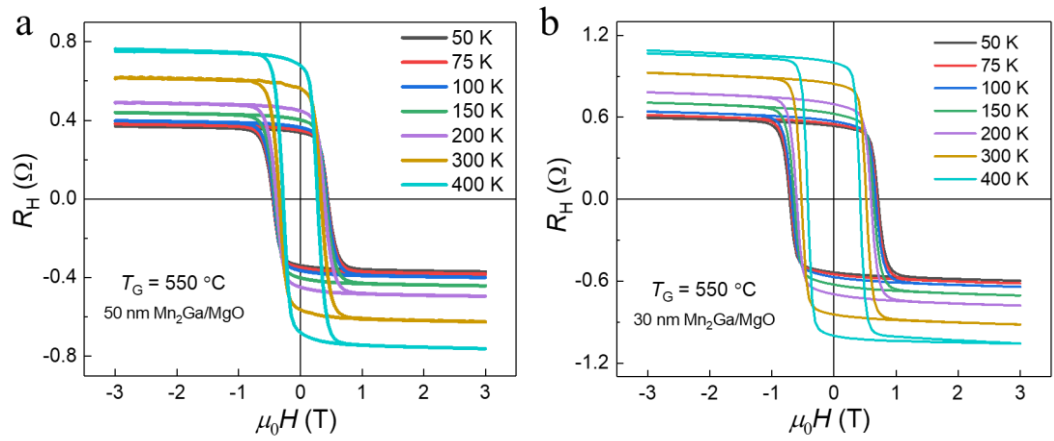

**Fig. 5 Hall effect of Mn₂Ga/MgO films with smaller thickness fabricated at $T_G$ = 550°C. a** 50 nm; **b** 30 nm

**Figure 6**

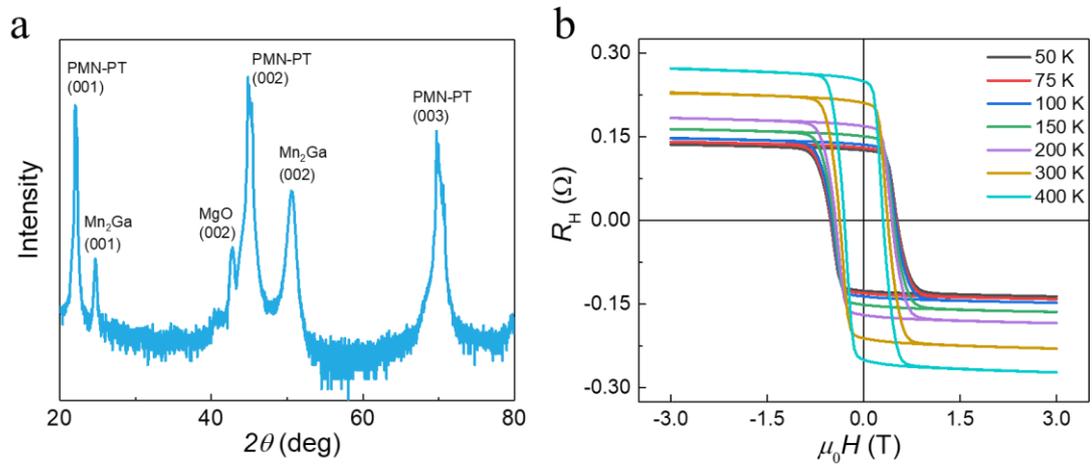

**Fig. 6 a** X-ray diffraction pattern of a Mn$_2$Ga/MgO/PMN-PT heterostructure; **b** Hall effect measurements of the Mn$_2$Ga/MgO/PMN-PT heterostructure at different temperatures ranging from 50 to 400 K

**Figure 7**

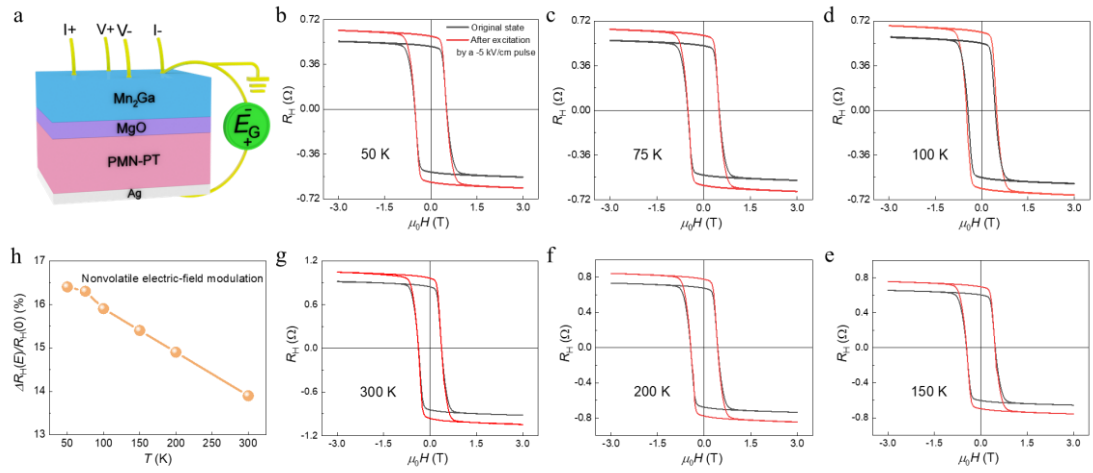

**Fig. 7 a** Schematic of the perpendicular electric-field modulation of the AHE for the Mn$_2$Ga(30 nm)/MgO(25 nm)/PMN-PT heterostructure; **b-g** AHE curves for the Mn$_2$Ga film before and after electric poling at various temperatures; **h** Relative zero-field Hall resistance modulation as a function of temperature

**Figure 8**

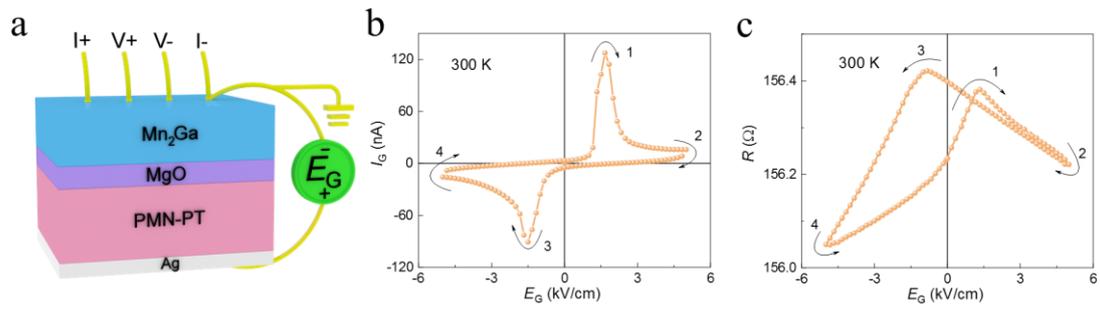

**Fig. 8 a** Schematic of the perpendicular electric-field modulation of the longitudinal resistance of $Mn_2Ga$ in a $Mn_2Ga$/MgO/PMN-PT heterostructure; **b** Room-temperature gating current as a function of perpendicular electric field; **c** Room-temperature longitudinal resistance of the $Mn_2Ga$ film versus electric field.

**Figure 9**

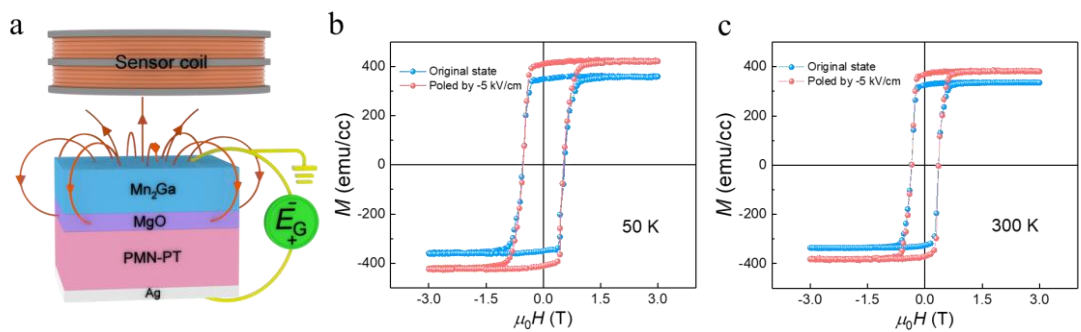

**Fig. 9 a** Schematic of the out-of-plane magnetization measurements upon electric-field poling of the ferroelectric PMN-PT substrate for the Mn$_2$Ga(30 nm)/MgO(25 nm)/PMN-PT heterostructure; **b** Out-of-plane magnetization of the Mn$_2$Ga film at 50 K before and after electric-field poling; **c** Out-of-plane magnetization of the Mn$_2$Ga film at 300 K before and after electric-field poling